\documentclass[a4paper,pdflatex]{article}
\usepackage{graphicx,amsmath,amsfonts}
\RequirePackage[english]{babel}
\usepackage[T2A]{fontenc}
\usepackage[cp1251]{inputenc}
\usepackage{enumerate}
\large

\title{\Large \bf Corrections to the Bethe formula for average ionization energy loss of relativistic charged particles
in solids. I. Mott's corrections}
\author{
O. Voskresenskaya}
\date{}

\begin{document}
\maketitle

\begin{center}
Joint Institute for Nuclear Research, Dubna, Moscow Region, 141980 Russia
\end{center}

\large

\begin{abstract}
Based on the proposed representation of the Mott corrections $\Delta_\textrm{M} $ to the Bethe stopping
formula in the form of rapidly convergent series of quantities bilinear in the Mott partial amplitudes,
the numerical calculations were performed for these corrections over the range of nuclear charge number
$ 4 \leq Z \leq 92 $ of the incident particles at various values of their relative velocity $\beta$.

\end{abstract}

\newpage
\large

\section{Introduction}

Stopping power (the average energy loss of a particle
per unit path length) is a necessary ingredient for many parts of nuclear and particle physics,
as well as for a wide variety of application areas within materials and surface science and engineering, 
micro and nano science and technology, 
radiation medicine and biology
\cite{1,2,3}.
Stopping power of a relativistic particle
is described by the relativistic version of the Bethe formula \cite{Bethe1, Bethe3}

\begin{equation}
\label{Bethe}
-\frac{d \bar E}{dx}=2\zeta\left[\ln\left(\frac{E_m}{I}\right)-\beta^2\right]\nonumber
\end{equation}
with
\begin{equation}
\zeta=K\left(\frac{Z}{\beta}\right)^2\frac{Z^\prime}{A},\quad K=2\pi N_A\frac{e^4}{mc^2},\quad
E_m\approx \frac{2m c^2\beta^2}{1-\beta^2}\,.
\end{equation}
Here, $x$ is the distance traveled by a particle,
$E_m$ denotes the maximum transferrable energy to an electron in a collision with the particle
of velocity $\beta c$, $m$ is the electron mass,
$Z^\prime$ and $A$  are the atomic number and the weight of an absorber, respectively, $N_A$ is the Avogadro number, and $I$
is its mean excitation potential. F. Bloch showed that  the mean
excitation potential of the atoms is approximately
given by $I=(10\,\mbox{eV})Z^\prime$ \cite{Bloch}. In this approximation the formula (1) is often called the `Bethe–Bloch formula' \cite{Ams}.


The importance of the Mott higher-order-correction term $\Phi_{Mott}/2$ \cite{Mott} to the formula (1)
\begin{eqnarray}
\label{BetheM}
-\frac{d\bar E}{dx}=2\zeta\left[\ln\left(\frac{E_m}{I}\right)-\beta^2 +\frac{\Phi_{Mott}}{2}\right]\,,\nonumber
\end{eqnarray}
\begin{equation}
\Phi_{Mott} = \frac{1}{\zeta}\Delta_{Mott}\!\!\left(\frac{d\bar E}{dx}\right),\quad
\Delta_{Mott}\!\!\left(\frac{d\bar E}{dx}\right)=N\!\!\int\!\!\left(\frac{d\sigma_{Mott}}{d\varepsilon}-
\frac{d\sigma_{Born}}{d\varepsilon}\right)\!\varepsilon d\varepsilon,
\label{Bethe3}
\end{equation}
was noted in various works (see, e.q., \cite{GSI1}). In the above formula
$\Delta_{Mott}$ denotes the Mott correction (MC),
$N=N_A Z^\prime/A$ is the number of target electrons per unit volume,  and $\sigma_{Born}$
represents the first-order Born approximation to the Mott exact cross section $\sigma_{Mott}$.

The expression for the MCs in (\ref{Bethe3}) is extremely inconvenient for practical application. In this regard, obtaining convenient and accurate representations for the  MCs become significant.
Ref. \cite{VSTT} gives an exact expression for the Mott correction  in the form of a
rather fast converging series of the quantities bilinear in the Mott partial amplitudes, which can be  quite simply calculated.

The aim of the presented work is to obtain numerical results for
the  Mott corrections  to the Bethe formula over a wide range of nuclear charge number
of the incident particles  at various values of their relative velocity $\beta$ based on results of \cite{VSTT}. 
The presented communication is organized as follows. Section 2 considers
an analytical result for the $\Delta_\textrm{Mott} $ correction
in the form of a quite rapidly converging series.
Section 3 gives the numerical results of their computation over the ranges $ 4 \leq Z \leq 92 $
and $ 0.75 \leq \beta \leq 0.95 $.
In Section 4 we briefly sum up our results and outline some prospects.

\section{Analytical result for Mott's correction to the Bethe stopping formula}

Switching in expression for $\Delta_{Mott}$ (3) from integration over the energy
 transferred to an target electron $\varepsilon$ to integration over the a center-of-mass scattering angle $\vartheta$, we rewrite this expression in the form
\begin{equation}
\Delta_{Mott}\!\!\left(\frac{d\bar E}{dx}\right)=N\frac{\pi}{m}\!\!\int_{\vartheta_0}^{\pi}\!\!\left[\omega_{Mott}(\vartheta)-
\omega_{Born}(\vartheta)\right]\!\sin^2(\vartheta/2)\sin\vartheta d\vartheta\,,
\label{M}
\end{equation}
where
\begin{equation}\omega_{Mott}(\vartheta)=\frac{\hbar^2}{4p^2\sin^2(\vartheta/2)}
\Bigl[\xi^{2}\vert F_{Mott}(\vartheta)\vert^{2}+\vert G_{Mott}(\vartheta)\vert^2\Bigl]\,,
\end{equation}
\begin{equation}
F_{Mott}(\vartheta)=\sum_{l}F_lP_l(x),~x=\cos \vartheta,\quad
G_{Mott}(\vartheta)=\sum_{l}G_lP_l(x)\, ,
\end{equation}
\begin{equation}
F_l=lC_l-(l+1)C_{l+1},\quad G_l=l^2C_l+(l+1)^2C_{l+1},
\quad C_l=\frac{\Gamma(\rho_l-i\nu)}{\Gamma(\rho_l+1+i\nu)}e^{i\pi(l-\rho_l)}\,,
\end{equation}
\begin{equation}
\label{sigma}
p=mc\frac{\beta}{\sqrt{1-\beta^2}},\quad\xi=\nu\sqrt{1-\beta^2},\quad
\rho_l=\sqrt{l^2-(Z\alpha)^2},\quad \nu=\frac{Z\alpha}{\beta}\,,
\end{equation}
\begin{equation}\omega_{Born}(\vartheta)=\frac{\nu^2}{\sin^4(\vartheta/2)}\left[1-
\beta^2\sin^2(\vartheta/2)\right]\,.
\end{equation}
Here, $P_l$  is the Legendre polynomial of order $l$
and $\Gamma(\mu)$ designates the Euler gamma function.

For what follows, instead of the original expression (6) for $G_{Mott}(\vartheta)$, it is convenient to employ a somewhat different expression  in terms of $F_{Mott}(\vartheta)$.

Writing
\begin{equation}
G_{Mott}(\vartheta)=\sum_{l}\left[l^2C_l+(l+1)^2C_{l+1}\right]P_l(x)\equiv
\sum_{l}(l+1)^2C_{l+1}\left[P_l(x)+P_{l+1}(x)\right]
\end{equation}
and taking account of the relation \cite{Ryzh}
\begin{equation}
(l+1)\left[P_l(x)+P_{l+1}(x)\right]=\cos(\vartheta/2)\left[P^{(1)}_{l+1}(x)+P^{(1)}_{l}(x)\right]\,,
\end{equation}
we obtain by an elementary calculation
\begin{equation}
G_{Mott}(\vartheta)=\cos(\vartheta/2)\sum_{l}\left[lC_l-(l+1)^2C_{l+1}\right]P^{(1)}_l(x)=
-\cos(\vartheta/2)F^\prime(\vartheta)\,
\end{equation}
and, in consequence,
\begin{equation}
\omega_{Mott}(\vartheta)=\frac{\hbar^2}{4p^2\sin^2(\vartheta/2)}
\Bigl[\xi^{2}\vert F_{Mott}(\vartheta)\vert^{2}+\vert F^{\prime}_{Mott}(\vartheta)\vert^2\Bigl]\,.
\end{equation}

Introducing
\begin{equation}
\tilde C_l=\frac{\Gamma(\rho_l-i\nu)}{\Gamma(\rho_l+1+i\nu)}\,,
\end{equation}
which are obtained from the $C_l$ by the substitution $\rho_l\to l$ and correspond the Sommerfeld--Moyer--Furry approximation \cite{SMF} in the theory of $eZ$ scattering, and the
corresponding quantities
\begin{equation}
\tilde F_l=l\tilde C_l-(l+1)\tilde C_{l+1}\,,
\end{equation}
as well as using the orthogonality relation for the Legendre function
\begin{equation}
\int\limits_{-1}^{1}P_{l}^{(m)}(\cos \vartheta)P_{l}^{(n)}(\cos \vartheta)\sin \vartheta d\vartheta=\frac{2}{2l+1}\,
\frac{(l+\vert m\vert) !}{(l-\vert m\vert)!}\delta_{ll'}\,,
\end{equation}
we can finally express the Mott correction as
\begin{equation}
\Delta_{Mott}\left(\frac{d\bar E}{dx}\right)=\frac{2\pi ZN_A}{mA}\sum_{l=0}^{L}\frac{\left[l(l+1)+\xi^2\right]}{2l+1}\left[\vert F_l\vert^2-
\vert \tilde F_l\vert^2\right]\,.
\label{Tarasov}
\end{equation}

It is easy to show that the terms of the series (\ref{Tarasov}) behave asymptotically 
as $l^{-2}$ and that the series converges absolutely.

\section{Numerical results for Mott's corrections to the Bethe formula}

The obtained result (16) allows us to reduce the calculation of the Mott corrections
$\Delta_{Mott}\left(d\bar E/dx\right)$ (3)
to the summation of a series consisting of quantities bilinear in the Mott partial amplitudes.
Table shows the computation results for the  corrections $\Phi_{Mott}/2$.
These results give the dependence of the $\Phi_{Mott}/2$  corrections on the nuclear charge number
of the incident particles and the values of their relative velocity.






\begin{center}
\noindent  {\bf Table 1.} $Z$ and $\beta$ dependence
of the Mott corrections $\Phi_{Mott}/2$  corrections to the Eq. (1).

\medskip
\begin{tabular}{lllll}
\hline\\[-.5cm]
Particle& $~~~~~~~Z~~~~~~~$&$\frac{1}{2}\Phi_{Mott}\vert_{\beta=0.75}$& $\frac{1}{2}\Phi_{Mott}\vert_{\beta=0.85}$& $\frac{1}{2}\Phi_{Mott}\vert_{\beta=0.95}$\\[.1cm]
\hline\\[-.5cm]
Be&04.000& 00.031& 00.035& 00.039\\
C &06.000& 00.041& 00.059& 00.070\\
Al&13.000& 00.118& 00.151& 00.160\\
Ti&22.000& 00.237& 00.248& 00.279\\
Fe&26.000& 00.283& 00.322& 00.353\\
Ni&28.000& 00.302& 00.339& 00.384\\
Mo&42.000& 00.516& 00.552& 00.631\\
Sn&50.000& 00.560& 00.670& 00.782\\
Ta&73.000& 00.981& 01.142& 01.302\\
W &74.000& 00.998& 01.158& 01.333\\
Pt&78.000& 01.060& 01.241& 01.431\\
Au&79.000& 01.076& 01.267& 01.450\\
Pb&82.000& 01.118& 01.335& 01.499\\
U &92.000& 01.251& 01.532& 01.769\\[.1cm]
\hline
\end{tabular}

\end{center}

\bigskip

It can be seen from  Table  that the Mott corrections
are becoming increasingly important
with the growth of $Z$ and $\beta$, and they are significant for nuclei of high charge number. 

\section{Summary and outlook}

\bigskip

\begin{itemize}

\item Based on the representation (16) of the Mott corrections
$\Delta_{Mott}\left(d\bar E/dx\right)$ to the Bethe formula (1) in the
form of quite rapidly converging series whose
terms are bilinear in the Mott partial amplitudes,
an algorithm is proposed to compute the $\Phi_{Mott}/2$ values
in the wide ranges of $Z$ and $\beta$.

\item This algorithm reduces the $\Phi_{Mott}/2$ computation  to
a summing the fast converging series (16) and
can be simply implemented using the numerical
summation methods of converging series for
a given level of precision.

\item Using the latter result,  the $\Phi_{Mott}/2$ corrections
to the Bethe stopping formula were calculated for
charged particles over the ranges  $ 4 \leq Z \leq 92 $
and $ 0.75 \leq \beta \leq 0.95 $.

\item It is shown that these corrections
are significant for nuclei of high $Z$.

\item It is of interest to find the total corrections of Mott and Bloch to the Bethe
stopping formula and to compare them with the Lindhard--S{\o}rensen corrections in the point nucleus approximation ($\gamma\leq 10$).

\item Comparison of the computational results with the available experimental data will be the subject of a further research.\\

\end{itemize}

\section*{Acknowledgments}

\bigskip

This work was financially supported by a grant from the Russian Foundation for Basic Research
(project nos. 17-01-00661-а) and partially supported
by a grant of the Plenipotentiary Representative of the Republic of Bulgaria at the JINR.


\end{document}